\documentclass[prl,twocolumn,showpacs,showkeys,amsmath,amssymb,floatfix]{revtex4}
\usepackage{graphicx}
\usepackage{bm}

\begin{document}

\title{Crossover from Attractive to Repulsive Casimir Forces and Vice
Versa} 
\author{Felix M. Schmidt} 
\author{H.~W. Diehl} 
\affiliation{%
  Fachbereich Physik, Universit{\"a}t Duisburg-Essen, 47048 Duisburg,
  Germany}

\date{\today}

\begin{abstract}
  Systems described by an $O(n)$ symmetrical $\phi^4$ Hamiltonian are
  considered in a $d$-dimensional film geometry at their bulk critical
  points. The critical Casimir forces 
  between the film's boundary planes $\mathfrak{B}_j,\,j=1,2$, are
  investigated as functions of film thickness $L$ for generic
  symmetry-preserving boundary conditions
  $\partial_n\bm{\phi}=\mathring{c}_j\bm{\phi}$. The $L$-dependent
  part of the reduced excess free energy per cross-sectional area 
  takes the scaling form $f_{\text{res}}\approx
  D(c_1L^{\Phi/\nu},c_2L^{\Phi/\nu})/L^{d-1}$ when $d<4$, where $c_i$
  are scaling fields associated with the 
  variables $\mathring{c}_i$, and $\Phi$ is a surface
  crossover exponent. Explicit two-loop renormalization group results for the
  function $D(\mathsf{c}_1,\mathsf{c}_2)$ at $d=4-\epsilon$ dimensions
  are presented. These show that (i) the Casimir force can
  have either sign, depending on $\mathsf{c}_1$ and $\mathsf{c}_2$, and
  (ii) for appropriate choices of the enhancements $\mathring{c}_j$,
  crossovers from attraction to repulsion and vice versa
  occur as $L$ increases.

\end{abstract}


\maketitle

Macroscopic bodies that are immersed in a medium frequently experience
long-range effective forces originating from fluctuations in the
medium. Such fluctuation-induced forces are ubiquitous in nature. A
well-known example is the Casimir force between two metallic
conducting plates caused by fluctuations of the electromagnetic field
\cite{Cas48}.  Other important examples are the Casimir forces caused by 
confined thermal fluctuations, either at critical points
\cite{FdG78,Kre94BDT00} or due to Goldstone modes \cite{KG99}.
Although predicted decades ago in a seminal paper by Fisher and de
Gennes \cite{FdG78}, such so-called ``thermodynamic Casimir forces''
were  verified experimentally in a clear manner only recently --- at first,
indirectly by the thinning of $^4$He wetting layers near the lambda
transition \cite{GC99} and subsequently by their direct observation
in binary fluid mixtures \cite{HGDB08}. Current Monte Carlo
simulations \cite{Huc07,VGMD07,HGDB08} were able to produce data in
conformity with these experiments.
 
Understanding fluctuation-induced forces is of great interest, for 
both technical and fundamental reasons. It has recently been realized
that quantum electromagnetic Casimir forces must be taken into
account when designing micromechanical devices \cite{CAKBC01}.
Analogous as well as thermodynamic Casimir forces are likely to be 
important also for microfluidic systems \cite{MC07,HGDB08}. From a general
vantage point, one of the most interesting aspects of fluctuation
induced forces is their \emph{universality}: They usually depend only
on gross features of the medium, the macroscopic bodies, and the geometry
but are independent of microscopic details. 

A much studied case is the Casimir force $\mathcal{F}_C$ between two
macroscopic parallel plates at a distance $L$, acting as boundaries of
the medium in the $z$~direction normal to the plates.  It is
frequently stated that, for a given medium and bulk dimension $d$, this
force --- and hence its sign --- depends on the boundary conditions $
\wp$ on both boundary planes. Both the QED Casimir force and the
thermodynamic Casimir force at a $d$-dimensional bulk critical point
decay as inverse powers of $L$, the former (in three dimension) as
$L^{-4}$, the latter as $L^{-d}$. Their strengths are commonly
characterized by dimensionless Casimir amplitudes $\Delta^{(\wp)}_C$,
which are believed to be universal, though boundary condition-dependent \cite{Kre94BDT00,KD9192a,DGS06,GD08}. For example, the critical
Casimir force (measured in temperature units $k_BT$ and per
cross-sectional area $A$) is conventionally written as
\begin{equation}
  \label{eq:FCcrit}
  \mathcal{F}_C=-(\partial/\partial L)
\Delta^{(\wp)}_CL^{-(d-1)}=(d-1)\, \Delta^{(\wp)}_C\,L^{-d}\;.
\end{equation}
If the boundary conditions are symmetric so that reflection positivity
holds, $\mathcal{F}_C$ is guaranteed to be attractive
\cite{Bac06KK06} (corresponding to $\Delta^{(\wp)}_C <0$); for
nonsymmetric boundary conditions, repulsive Casimir forces may occur
even for this simple slab geometry.

The aim of this Letter is to show that the above picture is
oversimplified.  Boundary conditions are \emph{scale-dependent
  properties}.  This entails that, even on length scales that are large compared
to microscopic distances, the strengths of the critical Casimir forces
cannot, in general, be characterized by constant universal amplitudes
$\Delta^{(\wp)}_C$.  Rather, the $\Delta^{(\wp)}_C$ get replaced by
effective scale- (i.e., $L$-) dependent amplitudes. As $L$ increases,
they can change considerably. Particularly interesting is that
\emph{even their signs may change} so that \emph{originally attractive
Casimir forces may turn repulsive} as $L$ increases and vice versa.
Focusing on the case of critical Casimir forces, we shall present
results for the associated scale-dependent amplitudes which
show that --- under appropriate conditions --- \emph{smooth crossovers
  from repulsive to attractive} as well as from \emph{attractive to
  repulsive Casimir forces} are possible. Such crossovers should be
accessible to experimental tests.

To become more specific, let us consider an $n$-component $\phi^4$
theory on a slab $\mathfrak{V}=\mathbb{R}^{d-1}\times [0,L]$ bounded
in the $z$ direction by a pair of planes $\mathfrak{B}_1$ at $z=0$ and
$\mathfrak{B}_2$ and $z=L$. For simplicity, we assume that these
planes do not give rise to interactions breaking the $O(n)$ symmetry
of the Hamiltonian $\mathcal{H}$. An appropriate choice then is
\cite{Die86a97,KD9192a,DGS06,GD08}
\begin{equation}
  \label{eq:Ham}
  \mathcal{H}=\int_{\mathfrak{V}}\bigg[\frac{1}{2}\,(\nabla\bm{\phi})^2  +\frac{\mathring{\tau}}{2}\,\phi^2+\frac{\mathring{u}}{4!}\,\phi^4\bigg] 
+\sum_{j=1}^2\int_{\mathfrak{B}_j}\,\frac{\mathring{c}_j}{2}\,\phi^2\;,
\end{equation}
where $\int_{\mathfrak{V}}$ and $\int_{\mathfrak{B}_j}$ are volume and
surface integrals, respectively. This Hamiltonian is well known from
the study of surface critical behavior. Let us recall some well-known
facts needed below \cite{Die86a97}. 

In Landau theory the Robin boundary conditions
\begin{equation}
  \label{eq:bc}
  \partial_n\bm{\phi}=\mathring{c}_j\bm{\phi} \quad\text{on }\mathfrak{B}_j
\end{equation}
result, where $\partial_n$ means a derivative along the inner normal.
For $\mathring{c}_j=\infty$ they reduce to Dirichlet and for
$\mathring{c}_j=0$ to Neumann boundary conditions.  The physical
significance of the interaction constants $\mathring{c}_j$ is to
account for local changes of the pair interactions near the planes
$\mathfrak{B}_j$. The larger the $\mathring{c}_j$, the stronger is the
order parameter $\bm{\phi}$ suppressed at $\mathfrak{B}_j$. In Landau
theory, $\mathring{c}_j=0$ corresponds to the special value
$\mathring{c}_j=\mathring{c}_{\text{sp}}$ at which the plane
$\mathfrak{B}_j$ becomes critical exactly at the bulk transition
temperature $T_{c,b}$. More precisely, when
$\mathring{c}_j<\mathring{c}_{\text{sp}}$, a continuous phase
transition from a disordered phase to a (bulk-disordered) phase with
long-range surface order at $\mathfrak{B}_j$ occurs in the
semi-infinite ($L=\infty$) system at a temperature
$T_{c,s}(\mathring{c}_j)>T_{c,b}$.  The special value
$\mathring{c}_{\text{sp}}$ of $\mathring{c}_j$ at which
$T_{c,s}(\mathring{c}_{\text{sp}})=T_{c,b}$ specifies a surface
multicritical point, called ``special'' on the bulk critical line
$\mathring{\tau}=\mathring{\tau}_{c,b}$. In Landau theory, both
$\mathring{\tau}_{c,b}$ and $\mathring{c}_{\text{sp}}$ vanish.

Beyond Landau theory, several important changes occur. First, the
boundary conditions~(\ref{eq:bc}) fluctuate --- they hold in an
operator sense, i.e.,\ inside of averages \cite{Die86a97}. Second,
provided the multicritical point exists, --- i.e.,\ when $d$ is
sufficiently large that long-range surface order is possible at
$T>T_{c,b}$ --- it gets shifted to nonzero values of
$\mathring{\tau}_{c,b}$ and $\mathring{c}_{\text{sp}}$, which depend
on microscopical details (lattice constant $a$ etc.).

Thus, for critical enhancement
$\mathring{c}_j=\mathring{c}_{\text{sp}}$, a Robin boundary
condition~\eqref{eq:bc} with a nonuniversal $\mathring{c}_{\text{sp}}$
rather than a Neumann boundary condition applies (on the mesoscopic
scale on which a continuum description is appropriate). This does not
automatically rule out the validity of a Neumann boundary condition in
the large-scale limit $z\to\infty$ with $a\ll z\lesssim \xi$ (where
$\xi$ is the bulk correlation length). The behavior of the order
parameter near $\mathfrak{B}_j$ follows from the boundary operator
expansion $\bm{\phi}(\bm{x})\approx C(\Delta
z)\,\bm{\phi}\rvert_{\mathfrak{B}_j}$, where
$\bm{\phi}\rvert_{\mathfrak{B}_j}$ is located on $\mathfrak{B}_j$ at a
distance $\Delta z$ from $\bm{x}$. The short-distance behavior
$C(\Delta z)\sim \lvert\Delta
z\rvert^{(\beta_1^{\text{sp}}-\beta)/\nu}$ is governed by the
difference of the scaling dimensions $\beta/\nu$ and
$\beta_1^{\text{sp}}/\nu$ of $\bm{\phi}$ and
$\bm{\phi}\rvert_{\mathfrak{B}_j}$, respectively. Only when their
difference vanishes does a Neumann boundary condition hold on large
length scales. While this is the case when $d$ exceeds the upper
critical dimension $d^*=4$ (since $\beta=\beta_1^{\text{sp}}=1/2$ in
Landau theory), it fails when $d<d^*$ because
$\beta>\beta_1^{\text{sp}}$. Thus, neither on mesoscopic nor on large
scales does a Neumann boundary condition hold at the special
transition when $d<d^*$.

Conversely, one may choose $\mathring{c}_j=0$, so that a Neumann
boundary condition holds on a mesoscopic scale, and inquire again into
the large-scale boundary conditions. Now $\mathring{c}_j=0$
translates into nonzero deviations $\delta\mathring{c}_j\equiv
\mathring{c}_j-\mathring{c}_{\text{sp}}$ from the multicritical point.
Hence the scaling fields $c_j\sim \delta\mathring{c}_j$ must vary
under changes $\mu\to\mu\ell$ of the momentum scale. They become
scale-dependent quantities $\bar{c}_j(\ell)$ that behave
$\sim\ell^{-\Phi/\nu}\,c_j$ near $c_j=0$, where $\Phi$ and $\nu$ are
the familiar surface crossover and bulk correlation length exponents,
respectively.  Depending on whether $c_j>0$ or $c_j<0$, they approach
the fixed-point values $c^*_{\text{ord}}=\infty$ and
$c^*_{\text{ex}}=-\infty$ at which the fixed points describing the
ordinary and extraordinary transitions of semi-infinite systems are
located. The short-distance behaviors of $\bm{\phi}$ near
$\mathfrak{B}_j$ are known in both cases. At the ordinary fixed point
$(c_j=+\infty$), one has $\bm{\phi}\sim
z^{(\beta_1^{\text{ord}}-\beta)/\nu}\partial_n\bm{\phi}
\rvert_{\mathfrak{B}_j}$, where $\beta_1^{\text{ord}}>\beta$ (for
$d<4$ \emph{and} in Landau, theory where $\beta_1^{\text{ord}}=1$); at
the extraordinary fixed point ($c_j=-\infty$), one has $\bm{\phi}\sim
\lvert\Delta z\rvert^{-\beta/\nu}$. Hence, whenever the initial
$c_j>0$, Dirichlet boundary conditions hold asymptotically on large
length scales at $\mathfrak{B}_j$. The upshot is that, for generic
values of $\mathring{c}_j\in(-\infty,\infty)$ and mesoscale boundary
conditions~\eqref{eq:bc}, the boundary conditions of the full
interacting theory will change under scale transformations, \emph{even
  in the Neumann case } $\mathring{c}_j=0$.

To elucidate the  consequences for the critical Casimir force, recall
that the reduced free energy of the slab per cross-sectional area $A\to\infty$
can be decomposed as
\begin{equation}
  \label{eq:Fdec}
  F/k_BTA=L\,f_b+f_s+f_{\text{res}}(L)
\end{equation}
into contributions from the bulk density $f_b$, the surface excess
density $f_s$, and an $L$-dependent residual part $f_{\text{res}}(L)$.
The behavior of these quantities at $T_{c,b}$ can be analyzed via
field-theoretic renormalization group (RG) methods. The RG equations
satisfied by $f_b$ and $f_s$ upon renormalization at $d<d^*$ are
inhomogeneous. However, the one of $f_{\text{res}}(L)$ is known to be
homogeneous \cite{Die86a97,KD9192a,DGS06,GD08}. Solving it at
$T_{c,b}$ yields the scaling form
\begin{equation}
  \label{eq:fresscf}
  f_{\text{res}}(L)/n\approx L^{-(d-1)}\,D(c_1L^{\Phi/\nu},c_2L^{\Phi/\nu})\;,
\end{equation}
where the $c_j$ now denote renormalized quantities
$c_j=\mu^{-1}Z_c^{-1}\delta\mathring{c}_j$ involving a familiar
renormalization factor $Z_c$ of Ref.~\cite{Die86a97}. The function
$D(\mathsf{c}_1,\mathsf{c}_2)$ is universal (up to nonuniversal metric
factors). It replaces the amplitude $\Delta^{(\wp)}_C/n$ in the first
form of Eq.~\eqref{eq:FCcrit}, while the critical Casimir force becomes
\begin{equation}
  \label{eq:Fc}
  \mathcal{F}_C/n\approx \mathcal{D}(c_1L^{\Phi/\nu},c_2L^{\Phi/\nu})\,L^{-d}
\end{equation}
with 
\begin{equation}
  \label{eq:Fscf}
  \mathcal{D}(\mathsf{c}_1,\mathsf{c}_2)=\big[d-1+(\Phi/\nu)\big(\mathsf{c}_1\partial_{\mathsf{c}_1}+ \mathsf{c}_2\partial_{\mathsf{c}_2}\big)\big]D(\mathsf{c}_1,\mathsf{c}_2)\,.
\end{equation}

We have computed the functions $D$ and $\mathcal{D}$ in
$d=4-\epsilon$ dimensions for general nonnegative values of
$\mathsf{c}_1$ and $\mathsf{c}_2$ to two-loop order, using $\epsilon$
as a small parameter. The required free-energy terms involve summations
over the spectrum $\{k_m^2\}$ of the operator $-\partial_z^2$ on
$[0,L]$. The discrete values $k_m$ are fixed by the boundary
conditions~\eqref{eq:bc} and depend on $\mathring{c}_1$, $\mathring{c}_2$, and $L$. We
evaluated such mode sums by means of complex integration, employing a
variant of Abel-Plana techniques that facilitated the separation of
bulk and surface terms \cite{rem:tbp}.

To present our results, we introduce the functions
\begin{equation}
  g_{\mathsf{c}_1,\mathsf{c}_2}(t)= \ln\left[1-
    \frac{(\mathsf{c}_1-t)(\mathsf{c}_2-t)}{(\mathsf{c}_1
      +t)(\mathsf{c}_2+t)}\,e^{-2t}\right]\;,  
\end{equation}
\begin{equation}
  D_0(\mathsf{c}_1,\mathsf{c}_2)=\frac{1}{4\pi^2}\int_0^\infty
  dt\,t^2\,g_{\mathsf{c}_1,\mathsf{c}_2}(t)\;,
\end{equation}
\begin{equation}
  J^{(\sigma)}_{\mathsf{c}_1,\mathsf{c}_2}=\int_0^\infty
  \frac{(-1)^\sigma\,
    t^{1+2\sigma}\,dt}{(t+\mathsf{c}_1)^2(t+\mathsf{c}_2)^2\,e^{2t}
    -(t^2-\mathsf{c}_1^2)(t^2-\mathsf{c}_2^2)},
\end{equation}
and the polynomials
\begin{eqnarray}
  P^{(0,0)}_{\mathsf{c}_1,\mathsf{c}_2}&=&
    2\mathsf{c}_1^3\mathsf{c}_2^3(\mathsf{c}_1+\mathsf{c}_2+\mathsf{c}_1\mathsf{c}_2),\nonumber\\
  P^{(1,1)}_{\mathsf{c}_1,\mathsf{c}_2}&=&2(\mathsf{c}_1^3+\mathsf{c}_2^3
  +2\mathsf{c}_1^2\mathsf{c}_2+2\mathsf{c}_1\mathsf{c}_2^2+(\mathsf{c}_1^2+\mathsf{c}_2^2)^2),\nonumber\\ 
P^{(2,2)}_{\mathsf{c}_1,\mathsf{c}_2}&=&2,\nonumber\\
P^{(1,0)}_{\mathsf{c}_1,\mathsf{c}_2}&=&2\mathsf{c}_1\mathsf{c}_2(\mathsf{c}_1^2+\mathsf{c}_2^2)
(\mathsf{c}_1+\mathsf{c}_2+\mathsf{c}_1\mathsf{c}_2) =P^{(0,1)}_{\mathsf{c}_1,\mathsf{c}_2},\nonumber\\
P^{(2,1)}_{\mathsf{c}_1,\mathsf{c}_2}&=&
2(\mathsf{c}_1+\mathsf{c}_2+\mathsf{c}_1^2+\mathsf{c}_2^2)=P^{(1,2)}_{\mathsf{c}_1,\mathsf{c}_2},
\nonumber\\
P^{(2,0)}_{\mathsf{c}_1,\mathsf{c}_2}&=&
2\mathsf{c}_1\mathsf{c}_2(\mathsf{c}_1+\mathsf{c}_2+\mathsf{c}_1\mathsf{c}_2)=P^{(0,2)}_{\mathsf{c}_1,\mathsf{c}_2}.
\end{eqnarray}
Then our result for $D$ can be written as
\begin{widetext}
\begin{eqnarray}\label{eq:Depsexp}
 D(\mathsf{c}_1,\mathsf{c}_2)&=&D_0(\mathsf{c}_1,\mathsf{c}_2)+\epsilon\bigg\{
  \Big(1-\frac{\gamma-\ln\pi}{2}\Big)\,D_0(\mathsf{c}_1,\mathsf{c}_2) 
-\frac{1}{4\pi^2}\int_0^\infty dt\,
g_{\mathsf{c}_1,\mathsf{c}_2}(t) \,t^2\ln t
\nonumber\\ &&\strut
 +\frac{n+2}{n+8}\bigg[\sum_{j=1}^2\Big(\frac{\gamma}{2}-1 +\ln
(2\mathsf{c}_j)\Big)\,\mathsf{c}_j\partial_{\mathsf{c}_j}
D_0(\mathsf{c}_1,\mathsf{c}_2)
+ \frac{1}{4\pi^2}\sum_{\sigma,\lambda=0}^2
P^{(\sigma,\lambda)}_{\mathsf{c}_1,\mathsf{c}_2}\,
J^{(\sigma)}_{\mathsf{c}_1,\mathsf{c}_2}\,
J^{(\lambda)}_{\mathsf{c}_1,\mathsf{c}_2}\bigg]\bigg\}+o(\epsilon) ,
\end{eqnarray}
\end{widetext}
where $\gamma$ is the Euler-Mascheroni constant.  

To check this result, one can set $(c_1,c_2)$ to $(\infty,\infty)$,
$(\infty,0)$, and $(0,0)$ and confirm by analytic calculation of the
required integrals that the respective series~\eqref{eq:Depsexp}
reduce to the $O(\epsilon)$ results of Ref.~\cite{KD9192a} for the
amplitudes $\Delta^{(\text{ord,ord})}_C/n$,
$\Delta^{(\text{ord,sp})}_C/n$, and $\Delta^{(\text{sp,sp})}_C/n$.
Note that the case $(c_1,c_2)=(0,0)$ is special: Unlike
$\Delta^{(\text{ord,ord})}_C$ and $\Delta^{(\text{ord,sp})}_C$, the
amplitude $D(0,0)$ does \emph{not} have an expansion in \emph{integer}
powers of $\epsilon$ but involves also \emph{half-integer powers}
$\epsilon^{k/2}$, with $k\ge 3$ (besides powers of $\ln\epsilon$ when
$k>3$) \cite{DGS06}.

\begin{figure}[htb]
  \centering
\includegraphics[width=0.8\columnwidth]{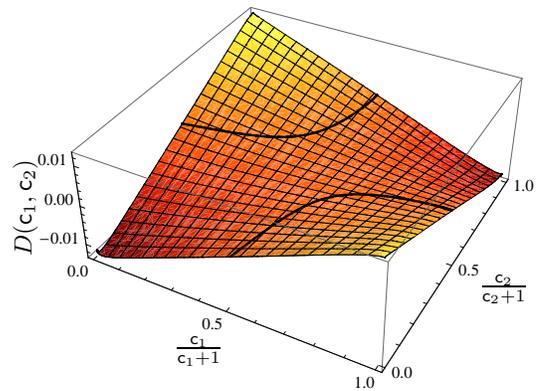}
  \caption{Scaling function $D(\mathsf{c}_1,\mathsf{c}_2)$ for $n=1$ and $d=3$. To cover
    the full domain $(0,\infty)^2$, we plotted $D(\mathsf{c}_1,\mathsf{c}_2)$ as a
    function of $\mathsf{c}_j/(1+\mathsf{c}_j)$. The zeros of $D$ are
    depicted as thick lines. \label{fig:D}}
\end{figure}

\begin{figure}[hbt]
  \centering
\includegraphics[width= 0.75\columnwidth]{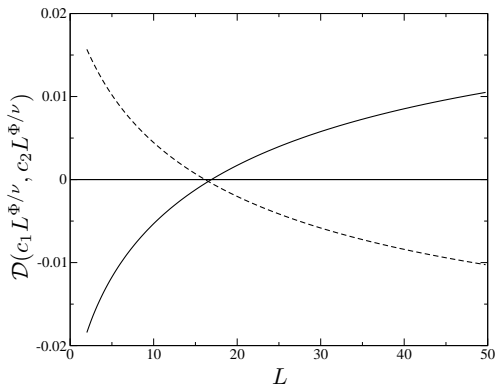}
  \caption{Scaled Casimir forces $\mathcal{F}_CL^d$ as  functions of
    $L$ for $(c_1,c_2)=(0,0.1)$ (solid line) and $(c_1,c_2)=(10,0.1)$
    (dashed line). In the first (second) case a crossover from
    attractive to repulsive (repulsive to attractive) Casimir forces
    occurs as $L\to \infty$. \label{fig:FC}}
\end{figure}

In Fig.~\ref{fig:D}, we show a plot of  $D(\mathsf{c}_1,\mathsf{c}_2)$ for the $d=3$ Ising
case $n=1$. It was obtained by numerical evaluation of the $O(\epsilon)$
result~\eqref{eq:Depsexp} at $\epsilon=n=1$.
As one sees, $D$ changes sign along certain paths. The same is true
for the scaling function~\eqref{eq:Fscf}, and hence for the critical
Casimir force $\mathcal{F}_C$. Moreover, such sign changes of
$\mathcal{F}_C$ occur upon increasing $L$ provided $(c_1,c_2)$ have
appropriate values. That crossovers from attractive to repulsive
Casimir forces and vice versa can occur is illustrated in
Fig.~\ref{fig:FC}.

To put these results in perspective, consider the case
$\mathring{c}_1=0$. Here a Neumann boundary condition holds at
$\mathfrak{B}_1$ for the regularized theory on the mesoscopic scale on
which the continuum description applies. One can derive this model
from a simple cubic lattice spin model whose ferromagnetic
nearest-neighbor bonds $J_{\bm{x}\bm{x}'}$ have strengths $J_j$
($j=1,2$) and $J$, depending on whether both sites $\bm{x}$ and $\bm{x}'$
belong to $\mathfrak{B}_j$ or at least one of them is not a boundary
site. From the known (approximate) relation between $\mathring{c}_j$
and $J_j/J$ \cite{Die86a97}, one sees that at $d=3$ the value
$(J_1/J)_0=5/4$ corresponds to $\mathring{c}_j=0$. This is less than
the value $\rho_{\text{sp}}= 1.500(4)$ at which the multicritical
point of the $d=3$ Ising model is located according to Monte Carlo
simulations \cite{LB90aRDW92PS98}. Thus, a mesoscopic Neumann boundary
condition at $\mathfrak{B}_j$ corresponds to a subcritical enhancement
$c_j>0$. As explained, a Dirichlet boundary condition applies in such
a situation on large length scales.

A similar crossover $\mathsf{c}_2=c_1L^{\Phi/\nu}\to \infty$ occurs
also for the choice $c_2=0.1$ made in Fig.~\ref{fig:FC}. In regimes 
where $\mathsf{c}_1$ and $\mathsf{c}_2$ are both small or both large,
$\mathcal{F}_C$ must be attractive. Yet in regimes where
$\mathsf{c}_1$ is sufficiently small while $\mathsf{c}_2$ is large,
$\mathcal{F}_C$ is repulsive. Depending on our choices $c_1=0$ and
$c_1=10$ ($\gg c_2$), crossovers from attractive to repulsive Casimir forces
and vice versa occur. 

These predictions should be testable by Monte Carlo
simulations for three-dimensional Ising models of the kind described
above and studied in Ref.~\cite{LB90aRDW92PS98}. Ideal experimental systems
to measure the calculated scaling function would satisfy three
criteria: (i) order-parameter dimension $n=1$; (ii) non-symmetry-breaking
boundaries; (iii) tunability of the effective boundary pair
interactions ($\mathring{c}_1$ and $\mathring{c}_2$). In the case of
$^4$He at the lambda transition, the boundaries do not break the
$O(2)$ symmetry. However, a two-component order parameter is involved,
a long-range ordered surface phase should not be possible at $d=3$,
and it is unclear to us how the parameters $\mathring{c}_j$ can be
varied. Experimental studies of binary liquid mixtures seem to us a more
promising alternative. For them, (i) is evidently satisfied. Since
walls usually favor one or the other component, the $\mathbb{Z}_2$
symmetry is broken by linear boundary terms
$-\int_{\mathfrak{B}_j}h_j\phi$, where each $h_j$ can have either
sign. It was demonstrated in Ref.~\cite{DPF95} that the values of these
fields $h_j$ can be changed by chemically modifying the surface. It is
also known that different signs of $h_j$ can be realized by proper
choices of the mixtures and substrates \cite{FYP05}. Hence it should be possible to
realize experimental setups with $h_1$ and $h_2$ small and of
opposite signs, or with $h_1\approx 0$ and $h_2$ large and of equal
signs. In both cases, sign-changing crossovers of the critical
Casimir forces may be expected as $L$ grows. The associated scaling
functions would need separate calculations. 

Partial support by DFG under grant Di 378/5 is gratefully
acknowledged.


\begin{thebibliography}{20}
\expandafter\ifx\csname natexlab\endcsname\relax\def\natexlab#1{#1}\fi
\expandafter\ifx\csname bibnamefont\endcsname\relax
  \def\bibnamefont#1{#1}\fi
\expandafter\ifx\csname bibfnamefont\endcsname\relax
  \def\bibfnamefont#1{#1}\fi
\expandafter\ifx\csname citenamefont\endcsname\relax
  \def\citenamefont#1{#1}\fi
\expandafter\ifx\csname url\endcsname\relax
  \def\url#1{\texttt{#1}}\fi
\expandafter\ifx\csname urlprefix\endcsname\relax\def\urlprefix{URL }\fi
\providecommand{\bibinfo}[2]{#2}
\providecommand{\eprint}[2][]{\url{#2}}

\bibitem[{\citenamefont{Casimir}(1948)}]{Cas48}
\bibinfo{author}{\bibfnamefont{H.~B.~G.} \bibnamefont{Casimir}},
  \bibinfo{journal}{Proc. K. Ned. Akad. Wet.} \textbf{\bibinfo{volume}{B51}},
  \bibinfo{pages}{793} (\bibinfo{year}{1948}).

\bibitem[{\citenamefont{Fisher and de~Gennes}(1978)}]{FdG78}
\bibinfo{author}{\bibfnamefont{M.~E.} \bibnamefont{Fisher}} \bibnamefont{and}
  \bibinfo{author}{\bibfnamefont{P.-G.} \bibnamefont{de~Gennes}},
  \bibinfo{journal}{C.\ R.\ S{\'e}ances.\ Acad.\ Sci.\ S{\'e}rie B}
  \textbf{\bibinfo{volume}{287}}, \bibinfo{pages}{207} (\bibinfo{year}{1978}).

\bibitem[{\citenamefont{Krech}(1994)}]{Kre94BDT00}
\bibinfo{note}{For background and lists of references, see}
\bibinfo{author}{\bibfnamefont{M.}~\bibnamefont{Krech}},
  \emph{\bibinfo{title}{{C}asimir Effect in Critical Systems}}
  (\bibinfo{publisher}{World Scientific}, \bibinfo{address}{Singapore},
  \bibinfo{year}{1994});
\bibinfo{author}{\bibfnamefont{J.~G.} \bibnamefont{Brankov}},
  \bibinfo{author}{\bibfnamefont{D.~M.} \bibnamefont{Dantchev}},
  \bibnamefont{and} \bibinfo{author}{\bibfnamefont{N.~S.}
  \bibnamefont{Tonchev}}, \emph{\bibinfo{title}{Theory of Critical Phenomena in
  Finite-Size Systems --- Scaling and Quantum Effects}}
  (\bibinfo{publisher}{World Scientific}, \bibinfo{address}{Singapore},
  \bibinfo{year}{2000}).

\bibitem[{\citenamefont{Kardar and Golestanian}(1999)}]{KG99}
\bibinfo{author}{\bibfnamefont{M.}~\bibnamefont{Kardar}} \bibnamefont{and}
  \bibinfo{author}{\bibfnamefont{R.}~\bibnamefont{Golestanian}},
  \bibinfo{journal}{Rev. Mod. Phys.} \textbf{\bibinfo{volume}{71}},
  \bibinfo{pages}{1233} (\bibinfo{year}{1999}).

\bibitem[{\citenamefont{Garcia and Chan}(1999)}]{GC99}
\bibinfo{author}{\bibfnamefont{R.}~\bibnamefont{Garcia}} \bibnamefont{and}
  \bibinfo{author}{\bibfnamefont{M.~H.~W.} \bibnamefont{Chan}},
  \bibinfo{journal}{Phys. Rev. Lett.} \textbf{\bibinfo{volume}{83}},
  \bibinfo{pages}{1187} (\bibinfo{year}{1999}).

\bibitem[{\citenamefont{Hertlein et~al.}(2008)\citenamefont{Hertlein, Gambassi,
  Dietrich, and Bechinger}}]{HGDB08}
\bibinfo{author}{\bibfnamefont{C.}~\bibnamefont{Hertlein}},
  \bibinfo{author}{\bibfnamefont{A.}~\bibnamefont{Gambassi}},
  \bibinfo{author}{\bibfnamefont{S.}~\bibnamefont{Dietrich}}, \bibnamefont{and}
  \bibinfo{author}{\bibfnamefont{C.}~\bibnamefont{Bechinger}},
  \bibinfo{journal}{Nature} \textbf{\bibinfo{volume}{451}},
  \bibinfo{pages}{172} (\bibinfo{year}{2008}).

\bibitem[{\citenamefont{Hucht}(2007)}]{Huc07}
\bibinfo{author}{\bibfnamefont{A.}~\bibnamefont{Hucht}},
  \bibinfo{journal}{Phys. Rev. Lett.} \textbf{\bibinfo{volume}{99}},
  \bibinfo{eid}{185301}  (\bibinfo{year}{2007}).

\bibitem[{\citenamefont{Vasilyev et~al.}(2007)\citenamefont{Vasilyev, Gambassi,
  Macio{\l}ek, and Dietrich}}]{VGMD07}
\bibinfo{author}{\bibfnamefont{O.}~\bibnamefont{Vasilyev}},
  \bibinfo{author}{\bibfnamefont{A.}~\bibnamefont{Gambassi}},
  \bibinfo{author}{\bibfnamefont{A.}~\bibnamefont{Macio{\l}ek}},
  \bibnamefont{and} \bibinfo{author}{\bibfnamefont{S.}~\bibnamefont{Dietrich}},
  \bibinfo{journal}{Europhys. Lett.} \textbf{\bibinfo{volume}{80}},
  \bibinfo{pages}{60009} (\bibinfo{year}{2007}).

\bibitem[{\citenamefont{Chan et~al.}(2001)\citenamefont{Chan, Aksyuk, Kleiman,
  Bishop, and Capasso}}]{CAKBC01}
\bibinfo{author}{\bibfnamefont{H.~B.} \bibnamefont{Chan}},
  \bibinfo{author}{\bibfnamefont{V.~A.} \bibnamefont{Aksyuk}},
  \bibinfo{author}{\bibfnamefont{R.~N.} \bibnamefont{Kleiman}},
  \bibinfo{author}{\bibfnamefont{D.~J.} \bibnamefont{Bishop}},
  \bibnamefont{and} \bibinfo{author}{\bibfnamefont{F.}~\bibnamefont{Capasso}},
  \bibinfo{journal}{Science} \textbf{\bibinfo{volume}{291}},
  \bibinfo{pages}{1941} (\bibinfo{year}{2001});
 \bibinfo{journal}{Phys. Rev. Lett.} \textbf{\bibinfo{volume}{87}},
  \bibinfo{pages}{211801} (\bibinfo{year}{2001});
\bibinfo{nore}{for earlier work on micromechanical systems, see}
\bibinfo{author}{\bibfnamefont{F.}~\bibnamefont{Serry}},
  \bibinfo{author}{\bibfnamefont{D.}~\bibnamefont{Walliser}}, \bibnamefont{and}
  \bibinfo{author}{\bibfnamefont{G.}~\bibnamefont{Maclay}},
  \bibinfo{journal}{J.\ Microelectromech.\ Syst.}
  \textbf{\bibinfo{volume}{4}}, \bibinfo{pages}{193}
  (\bibinfo{year}{1995}).

\bibitem[{\citenamefont{Munday and Capasso}(2007)}]{MC07}
\bibinfo{author}{\bibfnamefont{J.~N.} \bibnamefont{Munday}} \bibnamefont{and}
  \bibinfo{author}{\bibfnamefont{F.}~\bibnamefont{Capasso}},
  \bibinfo{journal}{Phys. Rev. A} \textbf{\bibinfo{volume}{75}},
  \bibinfo{eid}{060102(R)}  (\bibinfo{year}{2007});
\bibinfo{nore}{for further discussion of the agreement of this work's
  experimental data and Lifshitz theory, see}
\bibinfo{author}{\bibfnamefont{B.}~\bibnamefont{Geyer}},
  \bibinfo{author}{\bibfnamefont{G.~L.} \bibnamefont{Klimchitskaya}},
  \bibinfo{author}{\bibfnamefont{U.}~\bibnamefont{Mohideen}}, \bibnamefont{and}
  \bibinfo{author}{\bibfnamefont{V.~M.} \bibnamefont{Mostepanenko}},
  \bibinfo{journal}{\pra}
  \textbf{\bibinfo{volume}{77}}, \bibinfo{eid}{036102}
   (\bibinfo{year}{2008}) and
\bibinfo{author}{\bibfnamefont{J.~N.} \bibnamefont{Munday}} \bibnamefont{and}
  \bibinfo{author}{\bibfnamefont{F.}~\bibnamefont{Capasso}},
  \bibinfo{journal}{\pra}
  \textbf{\bibinfo{volume}{77}}, \bibinfo{eid}{036103}
   (\bibinfo{year}{2008}).

\bibitem[{\citenamefont{Krech and Dietrich}(1991)}]{KD9192a}
\bibinfo{author}{\bibfnamefont{M.}~\bibnamefont{Krech}} \bibnamefont{and}
  \bibinfo{author}{\bibfnamefont{S.}~\bibnamefont{Dietrich}},
  \bibinfo{journal}{Phys. Rev. Lett.} \textbf{\bibinfo{volume}{66}},
  \bibinfo{pages}{345} (\bibinfo{year}{1991});
  \bibinfo{journal}{Phys. Rev. A} \textbf{\bibinfo{volume}{46}},
  \bibinfo{pages}{1886} (\bibinfo{year}{1992}).



\bibitem[{\citenamefont{Diehl et~al.}(2006)\citenamefont{Diehl, Gr{\"u}neberg,
  and Shpot}}]{DGS06}
\bibinfo{author}{\bibfnamefont{H.~W.} \bibnamefont{Diehl}},
  \bibinfo{author}{\bibfnamefont{D.}~\bibnamefont{Gr{\"u}neberg}},
  \bibnamefont{and} \bibinfo{author}{\bibfnamefont{M.~A.} \bibnamefont{Shpot}},
  \bibinfo{journal}{Europhys. Lett.} \textbf{\bibinfo{volume}{75}},
  \bibinfo{pages}{241} (\bibinfo{year}{2006}),
  \bibinfo{note}{cond-mat/0605293}.

\bibitem[{\citenamefont{Gr{\"u}neberg and Diehl}(2008)}]{GD08}
\bibinfo{author}{\bibfnamefont{D.}~\bibnamefont{Gr{\"u}neberg}}
  \bibnamefont{and} \bibinfo{author}{\bibfnamefont{H.~W.} \bibnamefont{Diehl}},
  \bibinfo{journal}{Phys. Rev. B} \textbf{\bibinfo{volume}{77}},
  \bibinfo{eid}{115409} (\bibinfo{year}{2008}),
  \bibinfo{note}{{arXiv:0710.4436}}.

\bibitem[{\citenamefont{Bachas}(2006)}]{Bac06KK06}
\bibinfo{author}{\bibfnamefont{C.~P.} \bibnamefont{Bachas}},
 \bibinfo{journal}{J.\ Phys.\ A: Math. \& Gen.}
 \textbf{\bibinfo{volume}{40}},
 \bibinfo{pages}{9089}(\bibinfo{year}{2007}); 
\bibinfo{author}{\bibfnamefont{O.}~\bibnamefont{Kenneth}} \bibnamefont{and}
  \bibinfo{author}{\bibfnamefont{I.}~\bibnamefont{Klich}},
  \bibinfo{journal}{\prl} \textbf{\bibinfo{volume}{97}},
  \bibinfo{eid}{160401}  (\bibinfo{year}{2006}).

\bibitem[{\citenamefont{Diehl}(1986)}]{Die86a97}
\bibinfo{nore}{For background on boundary critical phenomena, see}
\bibinfo{author}{\bibfnamefont{H.~W.} \bibnamefont{Diehl}}, in
  \emph{\bibinfo{booktitle}{Phase Transitions and Critical Phenomena}}, edited
  by \bibinfo{editor}{\bibfnamefont{C.}~\bibnamefont{Domb}} \bibnamefont{and}
  \bibinfo{editor}{\bibfnamefont{J.~L.} \bibnamefont{Lebowitz}}
  (\bibinfo{publisher}{Academic}, \bibinfo{address}{London},
  \bibinfo{year}{1986}), vol.~\bibinfo{volume}{10}, pp.
  \bibinfo{pages}{75--267};
  \bibinfo{journal}{Int.\ J.\ Mod.\ Phys.\ B} \textbf{\bibinfo{volume}{11}},
  \bibinfo{pages}{3503} (\bibinfo{year}{1997}),
  \bibinfo{note}{cond-mat/9610143}.

\bibitem[{rem()}]{rem:tbp}
\bibinfo{note}{Details will be published elsewhere.}

\bibitem[{\citenamefont{Landau and Binder}(1990)}]{LB90aRDW92PS98}
\bibinfo{author}{\bibfnamefont{D.~P.} \bibnamefont{Landau}} \bibnamefont{and}
  \bibinfo{author}{\bibfnamefont{K.}~\bibnamefont{Binder}},
  \bibinfo{journal}{Phys.\ Rev.\ B} \textbf{\bibinfo{volume}{41}},
  \bibinfo{pages}{4633} (\bibinfo{year}{1990});
\bibinfo{author}{\bibfnamefont{C.}~\bibnamefont{Ruge}},
  \bibinfo{author}{\bibfnamefont{S.}~\bibnamefont{Dunkelmann}},
  \bibnamefont{and} \bibinfo{author}{\bibfnamefont{F.}~\bibnamefont{Wagner}},
  \bibinfo{journal}{Phys.\ Rev.\ Lett.} \textbf{\bibinfo{volume}{69}},
  \bibinfo{pages}{2465} (\bibinfo{year}{1992});
\bibinfo{author}{\bibfnamefont{M.}~\bibnamefont{Pleimling}} \bibnamefont{and}
  \bibinfo{author}{\bibfnamefont{W.}~\bibnamefont{Selke}},
  \bibinfo{journal}{Eur.\ Phys.\ J.\ B} \textbf{\bibinfo{volume}{1}},
  \bibinfo{pages}{385} (\bibinfo{year}{1998}).

\bibitem[{\citenamefont{Desai et~al.}(1995)\citenamefont{Desai, Peach, and
  Franck}}]{DPF95}
\bibinfo{author}{\bibfnamefont{N.~S.} \bibnamefont{Desai}},
  \bibinfo{author}{\bibfnamefont{S.}~\bibnamefont{Peach}}, \bibnamefont{and}
  \bibinfo{author}{\bibfnamefont{C.}~\bibnamefont{Franck}},
  \bibinfo{journal}{Phys. Rev. E} \textbf{\bibinfo{volume}{52}},
  \bibinfo{pages}{4129} (\bibinfo{year}{1995}).

\bibitem[{\citenamefont{Fukuto et~al.}(2005)\citenamefont{Fukuto, Yano, and
  Pershan}}]{FYP05}
\bibinfo{author}{\bibfnamefont{M.}~\bibnamefont{Fukuto}},
  \bibinfo{author}{\bibfnamefont{Y.~F.} \bibnamefont{Yano}}, \bibnamefont{and}
  \bibinfo{author}{\bibfnamefont{P.~S.} \bibnamefont{Pershan}},
  \bibinfo{journal}{Phys. Rev. Lett.} \textbf{\bibinfo{volume}{94}},
  \bibinfo{pages}{135702} (\bibinfo{year}{2005});
\bibinfo{author}{\bibfnamefont{S.}~\bibnamefont{Rafai}},
  \bibinfo{author}{\bibfnamefont{D.}~\bibnamefont{Bonn}}, \bibnamefont{and}
  \bibinfo{author}{\bibfnamefont{J.}~\bibnamefont{Meunier}},
  \bibinfo{journal}{Physica A} \textbf{\bibinfo{volume}{386}},
  \bibinfo{pages}{31} (\bibinfo{year}{2007}).
\end{thebibliography}
\end{document}